\documentstyle[aps,prb]{revtex}
\title{Stepwise quantum decay of self-localized solitons}
\author{V.Hizhnyakov and D.Nevedrov}
\address{Institute of Theoretical Physics, University of Tartu,
T\"{a}he 4, EE2400 Tartu, Estonia. \\
Institute of Physics, Riia 142, EE2400 Tartu, Estonia. \\
e-mail: hizh@park.tartu.ee}
\begin{document}
\maketitle
\begin{abstract}
The two-phonon decay of self-localized soliton in a one-di\-men\-sional
monatomic anharmonic lattice caused by cubic anharmonicity is considered.
It is shown that the decay takes place with emission of phonon bursts.
The average rate of emission of
phonons is of the order of vibrational quantum per vibrational period.
Characteristic time of the relaxation is determined by the quantum
anharmonicity parameter; this time may vary from a few (quantum lattices,
large anharmonicity)
to thousands (ordinary lattices, small anharmonicity) of vibrational periods.
\end{abstract}
\vspace{3mm}
\hspace{2.2mm}{\footnotesize PACS numbers: 63.20Kr, 63.20Pw, 63.20Ry}

The dynamics of strong nonlinear excitations in polymers and
quasi-one-di\-men\-sio\-nal biomolecular chains is an active research field.
Theoretical studies of anharmonic perfect lattices have shown the existence of
localized vibrations (self-localized solitons, SLSs) with frequencies above
the phonon band or in the gap of the phonon spectrum
(see \cite{ov,kosevich,dolgov,sivtak,page,burlakov,zavt,kiselev,sanpage,wagner}
and references
therein). SLSs are soliton-like excitations in discrete lattices, and are thus
closely related to ordinary solitons. The existence of stationary and moving
SLSs was derived in the frame of classical mechanics.
Until now there was a little discussion
about the influence of quantum (and thermal) fluctuations on stability of the
SLS. An exception is the paper of Ovchinnikov \cite{ov} who argues that the
decay of the SLS caused by these fluctuations diminishes with the increase of
the mode amplitude. This statement, however, is based on a perturbational
consideration which can not be applied to the description of evolution of
vibrations with a strong amplitude.

Recently we showed \cite{procest,hizhrev,hizhnev} that the perturbational
treatment of the effect of quantum and thermal fluctuations on a local mode
associated with a defect atom in a lattice fails for the case when the
amplitude of vibration is large. In fact, the two-phonon
damping of the local mode, caused by cubic anharmonicity,
behaves dramatically with the change of
the amplitude: at definite ``critical'' amplitudes relaxation jumps take place
being accompanied by a generation of phonon bursts. This effect indicates that
the quantum and thermal fluctuations may dominate in the dynamics of strong
vibrations with the energy of the mode being within a specific range. The
strong field of the local vibration causes the transformation
of phonon operators and the increase of the number of phonons in time
\cite{procest,hizhrev}. This
mechanism of phonon generation by a local vibration is analogous to the
mechanism of black hole radiation \cite{birel,grib,hawking}.

In this com\-munic\-ation we ex\-tend the theory \cite{procest,hizhrev,hizhnev} to
the SLS in a mon\-at\-omic one-di\-men\-sio\-nal lattice (chain) with
cubic and hard quartic anhar\-monicity. This case is of
special interest since the monatomic chain with both anharmonicities is the
simplest model for the investigation of the quantum relaxation of SLS
in a perfect lattice. Hard quartic anharmonicity is a prerequisite of
existence of the SLS, while the cubic anharmonicity stands for the
two-phonon decay of the SLS. We note that SLSs in this model
are stable in classical limit, since all harmonics  of the mode are out
of resonance with lattice phonons \cite{dolgov}. The odd SLS is examined and
analytical nonperturbative solution of the problem and results of
numerical calculations are presented.

The potential energy operator in a monatomic one-di\-men\-sio\-nal lattice,
which includes linear, and the first two nonlinear terms and takes account of
the nearest neighbour interaction, has the following form:
\begin{equation}
\hat{V} =
\sum_{n}{\sum_{r=2}^{4}{\frac{K_r}{r}(\hat{U}_{n+1} - \hat{U}_{n})^{r}}}\,,
\end{equation}
where $\hat{U}_{r}$ is the operator of the longitudinal displacement of the
$n$-th atom from its equilibrium position, $K_r$ are harmonic ($r=2$) and
anharmonic (cubic: $r=3$, quartic: $r=4$) springs. The operators $\hat{U}_{n}$
satisfy the following equations of motion:
\begin{equation}
\frac{\partial^2 \hat{U}_n}{\partial t^2} =
\sum_{r=2}^{4} \bar{K}_r {\Big [}(\hat{U}_{n+1}-\hat{U}_{n})^{r-1}
- (\hat{U}_{n} - \hat{U}_{n-1})^{r-1}{\Big ]}\,,
\end{equation}
$\bar{K}_r = {K_{r}}/{M}$,
$M$ is the mass of an atom. We suppose that an SLS with the frequency
$\omega_{l} < 2\omega_{D}$ is excited at the time $t=0$ at the sites $n=0$ and
its neighbours ($\omega_{D}$ is the maximum harmonic frequency,
$\omega_{D} = 2\sqrt{\bar{K}_{2}}$). Anharmonic interactions are supposed to
be weak, satisfying the condition $a_{0}^{(0)} \ll A_{0}$, where
${a}_{0}^{(0)} \sim \sqrt{\hbar / (2\omega_{D} M)}$ is the amplitude of
zero-point vibrations, $A_0 \sim \sqrt{K_2/K_4}$ is the amplitude of the
self-localized vibration (from physical reasons the value of $A_0$ should
not exceed the value of the lattice constant $d$,
i.e. $K_{4} > K_{2}/d^{2}$).
The condition $a_{0}^{(0)} \ll A_{0}$ means that the characteristic energy of the SLS, being of
the order of $K_{2}^{2}/K_{4}$ is much larger than the characteristic
vibrational quantum $\hbar \omega_{D}$:
\begin{equation}
{\cal K} = \frac{{K_{2}^{2}}/{K_{4}}}{\hbar \omega_{D}}
= \frac{K_{2}^{3/2}M^{1/2}}{2\hbar K_{4}} \gg 1 \,.
\end{equation}
Note that the reversed dimen\-sion\-less parameter ${\cal K}^{-1}$
characterizes the degree
of the quantum anharmonicity of the lattice: it increases as the anharmonic
term $K_{4}$ increases and as the mass of the atoms $M$ decreases.
In quantum crystals {\rm He} and
{\rm Ne} ${\cal K}$ is of the order of 1, while in ordinary crystals
(characterized by small amplitude of zero-point vibrations as compared to the
lattice constant) it can reach many hundreds. In the problem under
investigation ${\cal K}$ plays an essential role  determining the time-scale
of the energy relaxation (see below).

In order to take account of the SLS decay we introduce the operators
$\hat{U}_{n}$ in the form
\begin{equation}
\hat{U}_{n} = u_{n}(t) + \xi_{n} + \hat{q}_{n}\,,
\end{equation}
where the "classical" displacements $u_n$ are supposed to be nearly periodic
functions:
\mbox{$u_n(t) \simeq A_n(t) \cos{\omega_{l} t}$,} $A_{n}(t)$ and $\xi_{n}(t)$
are, slowly changing with time, amplitudes and shifts satisfying the equations
\begin{eqnarray}
-\omega_{l}^{2} A_{n} &  = & \bar{K}_{2}(\bar{A}_{n+1} - \bar{A}_{n}) +
2\bar{K}_{3}(\bar{A}_{n+1} \bar{\xi}_{n+1} - \bar{A}_{n} \bar{\xi}_{n}) +
\nonumber \\
& + & 3\bar{K}_{4} {\Big (}\frac{\bar{A}_{n+1}^{3}}{4} + \bar{A}_{n+1}
\xi _{n+1}^{2} -
\frac{\bar{A}_{n}^{3}}{4} - \bar{A}_{n} \bar{\xi}_{n}^{2}{\Big )}\,,
\label{eq:one}
\end{eqnarray}
\vspace{-5mm}
\begin{eqnarray}
0 & = & \bar{K}_{2} (\bar{\xi}_{n+1} - \bar{\xi}_{n}) +
\bar{K}_{3}{\Big (}\frac{\bar{A}_{n+1}^{2}}{2}+\bar{\xi}_{n+1}^{2}-
\frac{\bar{A}_{n}^{2}}{2}
-\bar{\xi}_{n}^{2}{\Big )} + \nonumber \\
& + & \bar{K}_{4} {\Big (}\frac{3\bar{A}_{n+1}^{2}\bar{\xi}_{n+1}}{2}
+\bar{\xi}_{n+1}^{3}-\frac{3\bar{A}_{n}^{2}\bar{\xi}_{n}}{2}-
\bar{\xi}_{n}^{2}{\Big )} \,,
\label{eq:two}
\end{eqnarray}
where we keep just the $\cos{\omega t}$ term neglecting
$\sim \cos{3\omega_{l} t}$,
$\cos{5\omega_{l} t}$, $\ldots$
\cite{dolgov,sivtak,page} and introduce the new variables: $\bar{A}_{n} =
A_{n} - A_{n-1}$, $\bar{\xi}_{n} = \xi_{n} - \xi_{n-1}$.
Variations of the $A_n$ and $\xi_{n}$ in time, due to quantum fluctuations,
are described by means of time-dependent operators $\hat{q}_{n}(t)$, which
satisfy the equations
\begin{equation}
\frac{d^2\hat{q}_n}{dt^2} = W_{n+1}(t) (\hat{q}_{n+1} - \hat{q}_{n})
- W_{n}(t) (\hat{q}_n - \hat{q}_{n-1})\,.
\end{equation}
Here $W_{n}(t) = \bar{K}_{2} + k_{n} + 2w_{n} \cos{\omega_{l}t}$, where
$k_{n} = 2\bar{K}_{3} \bar{\xi}_{n}+3\bar{K}_{4} {\Big (}{\bar{A}_{n}^{2}}/{2}
+ \bar{\xi}_{n}^{2}{\Big )}$ and $w_{n} = \bar{K}_{3} \bar{A}_{n} + 3
\bar{K}_{4} \bar{A}_{n} \bar{\xi}_{n}$ determine the change of springs caused
by the SLS. The terms $\sim K_{3}(\hat{q}_{n} -\hat{q}_{n-1})^{2}$,
$K_{4} \bar{A}_{n} (\hat{q}_{n} - \hat{q}_{n-1})^{2}$,
$K_{4} \bar{\xi}_{n} (\hat{q}_{n} - \hat{q}_{n-1})^{2}$ and
$K_{4}(\hat{q}_{n}-\hat{q}_{n-1}
)^{3}$ are supposed to be small and they were neglected. It is important to
take these terms into account when $\omega_{l} > 2\omega_{D}$, i.e. when the
two-phonon decay under consideration is forbidden by the energy conservation
law.

The phonon hamiltonian is defined as $\hat{H}_{ph}(t)=\hat{T}(t)+\hat{W}(t)$,
where $\hat{T} = \frac{1}{2}\sum_{n}\dot{\hat{q}}_{n}^{2}$ and
$\hat{W} = \hat{W}_{0} + \hat{W}_{1} + \hat{W}_{t}$ are the operators
of kinetic and potential energy, $\hat{W}_{0} = K_{2}/2\sum_{n}
(\hat{q}_{n}-\hat{q}_{n-1})^{2}$, $\hat{W}_{1} = 1/2\sum_{n}k_{n}(\hat{q}_{n}
- \hat{q}_{n-1})^{2}$, $\hat{W}_{t} = \cos{\omega_{l} t}\sum_{n}
w_{n}(\hat{q}_{n}-\hat{q}_{n-1})^{2}$; $\hbar =1$. $\hat{W}_{1}$ describes the
stationary perturbation of the phonon subsystem by the SLS, while
$\hat{W}_{t}$ takes account of the oscillatory time-dependence of the
springs induced by the SLS. This is the time-dependence of $\hat{H}(t)$
which causes the phonon emission by the SLS (phonons are generated by
pairs \cite{procest,hizhrev}).

Introducing properly chosen configurational
coordinates $\hat{X}_i = \sum_n S_{in}\hat{q}_{n}$ and $\hat{Q}_{m} =
\sum_{n}s_{mn}\hat{q}_{n}$, one can diagonalize $\hat{W}_{1}$ and
$\hat{W}_{t}$:
$\hat{W}_{1}  = \frac{1}{2} \sum_{i}\eta_{i} \hat{X}_{i}^{2}$,
$\hat{W}_t= \cos{\omega_{l} t} \sum_{m}v_{m} \hat{Q}_{m}^{2}$.
Both $\hat{W}_{1}$ and $\hat{W}_{t}$ have one zero eigenvalue; all other
$\eta_{i}$ are positive, while $v_{m}$ are sign-alternating, being symmetric
with respect to the sign change. The mode with zero eigenvalue is totally
symmetric and therefore does not enter into $\hat{q}_{n} - \hat{q}_{n-1}$.

The Hamiltonian $\hat{H}_{ph}$ is diagonalized by introducing time-dependent
operators as follows. First the time-in\-de\-pen\-dent Hamiltonian
$\hat{H} = \hat{H}_{ph}^{(0)} + \hat{W}_{0}$ is diagonalized by applying the
standard methods of local dynamics \cite{marad,maradu} ($\hat{H}_{ph}^{(0)}$
is the Hamiltonian of the chain in harmonic approximation). Then the
Hamiltonian $\hat{H}_{ph}(t)$ is diagonalized by the method presented in
\cite{procest,hizhrev,hizh}. The time-dependent phonon Hamiltonian in the
diagonal representation allows one to find energy of phonons $E_{ph}(t)$ and
the rate of phonon generation. The latter equals to \cite{procest,hizhrev}
\begin{equation}
\frac{dE_{ph}(t)}{dt} = - \frac{dE_{l}(t)}{dt} = I(t)\,.
\label{eq:deph}
\end{equation}
Here $E_{l} \simeq \sum_{n}{\Big (}\omega_{l}^{2} M \frac{A_{n}^{2}}{2} +
\sum_{r}K_{r}\frac{\xi_{n}^{r}}{r}{\Big )}$ is the energy of the SLS,
\begin{equation}
I =  \int_{\omega_{l}-\omega_{D}}^{\omega_{D}} \!\!\! d\omega
\sum_{m} \frac{\pi \omega_{l} v_{m}^2}{8}
\frac{\rho_{m}(\omega) \rho_{m}(\omega_{l} - \omega)
(1+2n(\omega))}{|1-v_{m}^{2}\tilde{G}_{m}(\omega)|^2}
\label{eq:power}
\end{equation}
denotes the intensity of the emission of phonons,
$\rho_{m} = \frac{2}{\pi} \Im (G_{m}(\omega))$, the temperature factor is
included by the multiplier $(1+2n(\omega))$, $\tilde{G}_{m}(\omega) =
G_{m}(\omega) G_{m}(\omega-\omega_{l})$,
\begin{equation}
G_{m}(\omega) = -i \int_{0}^{\infty} \!\! dt e^{i\omega t - \epsilon t}
\langle [\hat{Q}_{m}(t), \hat{Q}_{m}(0)]\rangle \,,
\quad
\epsilon \rightarrow 0\,,
\end{equation}
is the  Green's function of the mode $m$, $\hat{Q}_{m}(t)
\equiv e^{i\hat{H} t} \hat{Q}_{m} e^{-i\hat{H}t}$.

Use of the diagonal representation for the operator $\hat{W}_{1}$ allows one to
easily express the Green's functions $G_{m}$ via the one-cite Green's function
of the perfect lattice \cite{econ}
$G_{0}^{(0)} = i/(\omega \sqrt{\omega_{D}^{2} - \omega^{2}})$ as follows
\begin{equation}
G_{m} = G_{0}^{(0)}\!\sum_{nn'ii'}\!\frac{\sigma_{im}
\sigma_{i'm} S_{in} S_{i'n'} \cos{(\kappa (n\!-\!n'))}}
{{\big (}1-\eta_{i} G_{i}^{(0)}{\big )}
{\big (}1-\eta_{i'} G_{i'}^{(0)}{\big )}}
{\Big [}1 - \delta_{ii'}\eta_{i} G_{i}^{(0)}{\Big ]}\,.
\label{eq:gm}
\end{equation}
Here $G_{i}^{(0)} = G_{0}^{(0)} \sum_{nn'}S_{in}S_{in'} \cos{(k(n-n'))}$,
$\sigma_{mi} = \sum_{n} s_{mn}S_{in}$, $\cos{\kappa}
= 1 - 2\omega^2/\omega_{D}^2$, $\omega^2 < \omega_{D}^2$.

In the limit of small $|v_m|$ one can neglect the term
$v_{m}^2 \tilde{G}_m(\omega)$ in the denominator of formula (\ref{eq:power}).
In this limit (9) coincides with the corresponding formula of the
perturbation theory \cite{klemens}. For
realistic $K_{3}$ and the considered values of $\omega_{l}$ between
$1.3\omega_{D}$ and $2 \omega_{D}$ the term $v_{m}^{2}\tilde{G}_{m}$ is
not small. At some $\omega_{l} = \omega_{k}$ and $\omega$ both
$1-v_{m}^{2}\Re (\tilde{G}_{m}(\omega))$ and $\Im (\tilde{G}_{m}(\omega))$ turn
to zero and the perturbation theory fails. Near such $\omega_{k}$
$I \sim |\omega_{l}-\omega_{k}|^{-1}\sim |t\!-\!t_{k}|^{-1/2}$, i.e a sharp
burst of phonons is generated causing a relaxation jump ($t_{k}$ corresponds
to the time moment when $\omega_{l} = \omega_{k}$).

Note that $v_{m}^2$ is not a linear function of the energy of the SLS.
Therefore the relaxation of the SLS is always nonexponential, including the
case when perturbation theory is applied. In the latter case
$E_{l}(t) \sim (t_{l} - t)^{\alpha}$, where $\alpha \sim 0.75 \ldots 0.83$,
$t_{l}$ is the lifetime of the SLS which is finite in this approximation.

As an example the quantum decay of an odd SLS is examined below. The
properties of the SLS depend essentially on the dimen\-sion\-less parameter
$\delta = \sqrt{K_{3}^{2}/(K_{2} K_{4})}$. For realistic one-well
pair-potentials this parameter has the value between $1$ and $\sqrt{32/9}$.
For {\rm Ar-Ar} and {\rm Kr-Kr} potentials $\delta = 1.37$, for {\rm K-Br}
potential $\delta=1.31$. For such values of $\delta$ the odd SLS is well
localized if $\omega_{l} \stackrel{>}{\sim} 1.3\omega_{D}$. In this case
SLS is localized on three central atoms:
$A_{1} = A_{-1} \simeq -A_{0}/2$, $|A_{n}| \stackrel{<}{\sim} 0.05 |A_{0}|$,
$n \geq 2$; $\xi_{n} = -\xi_{-n} \simeq \xi$, $n \geq 1$, $\xi_{0} = 0$
\cite{dolgov,sivtak}. This allows one to take account of only $A_{0}$,
$A_{1}$ and
$\bar{\xi}_{1} = -\bar{\xi}_{-1} \simeq \xi$. The
parameters $A_{0}$ and $\xi$ of the classical SLS problem in this
approximation are determined by the equations (\ref{eq:one})
($n=0$) and (\ref{eq:two}) ($n=1$):
\begin{equation}
\omega_{l}^{2} = 3\bar{K}_{2} +  6\bar{K}_{3} \xi + \frac{81}{16}
\bar{K}_{4} A_{0}^{2} + 9\bar{K}_{4} \xi^{2}\,,
\label{eq:wl}
\end{equation}
\begin{equation}
0 = \bar{K}_{2}\xi + \bar{K}_{3} (A_{0}^{2} + \xi^{2}) + \bar{K}_{4}\xi
{\Big (}\frac{27}{8}A_{0}^{2} + \xi^{2}{\Big )}\,.
\label{eq:zero}
\end{equation}
Introducing dimen\-sion\-less amplitude $a = \sqrt{\bar{K}_{4}A_{0}^{2}/
\bar{K}_{2}}$ and shift $z = \xi\sqrt{\bar{K}_{4}/\bar{K}_{2}}$, one gets
\begin{equation}
a^2 = -\frac{z(z^2 +z\delta +1)}{\delta + \frac{27}{8}z}, \nonumber\\
\end{equation}
where z is determined by the real solutions of the equation
\begin{equation}
\omega_{l}^2 = \frac{3}{4} + \frac{81}{64}\frac{z(5z^2 + \frac{43}{9}z\delta +
\frac{32}{27}\delta^2 - 1)}{\delta + \frac{27}{8}z} \nonumber\\
\end{equation}
(here and below the units $\omega_{D} = 1$ are used). This is a cubic equation
for $z$ which for considered $\omega_{l}$ and $\delta$ has only one solution
with $a^{2}>0$. This solution determines the parameters of the SLS in
the classical limit.

In the approximation considered, the SLS causes changes of
only in central and next to  central springs:
\begin{eqnarray}
k_{0} & = &k_{1} \!=\! (16z\delta \!+\! 24z^{2}
\!+ \!27a^2)/32\,,
\quad
k_{2} \! =\! k_{-1} \!=\! {3a^2}/{32}\,, \nonumber \\
w_{0} & = & -w_{1} = {a(\delta + 3z)}/{8}\,,
\quad
w_{2}  = -w_{-1} = {a\delta}/{8}\,. \nonumber
\end{eqnarray}
In this case $\hat{W}_{1}$ and $\hat{W}_{t}$ can be
diagonalized analytically. The eigenvalues $\eta_i$ and $v_{m}$ and the
components of the eigenvectors are the following: $\eta_{1} = 0$,
\vspace{-3mm}
\begin{eqnarray}
&\eta_{2,3} &={\Big (}k_2+\frac{3k_{0}}{2} {\Big )} \pm {\Big (}(k_{2}-
\frac{3k_{0}}{2})^{2}
+k_{2}{\Big )}^{1/2}\,,
\nonumber \\
&\eta_{4,5} & = {\Big (}k_{2}+\frac{k_{0}}{2}{\Big )} \pm {\Big (}(k_{2}-
\frac{k_{0}}{2})^2-k_{2}{\Big )}^{1/2}\,,
\nonumber \\
& v_{3} & =  0\,,
\quad
v_{1,2,4,5} = \pm \frac{1}{2}{\big (}8w_{2}^{2}+6w_{1}^{2}+4w_{1}w_{2} \pm
\nonumber \\
&  & 2(16w_{2}^{4}+8w_{1}^{2}w_{2}^{2}+16w_{2}^{3}w_{1}+9w_{1}^{4}+12w_{1}^
{3}w_{2})
^{\frac{1}{2}}{\big )}^{\frac{1}{2}} \nonumber
\end{eqnarray}
($++$ corresponds to $v_{1}$, $-+$ to $v_{2}$, $+-$ to $v_{4}$
and $--$ to $v_{5}$);
\begin{eqnarray}
&& S_{kn} =  \frac{\tilde{S}_{kn}}{\sqrt{\sum_{n}\tilde{S}^{2}_{kn}}}\,,
\quad
\tilde{S}_{1n}  = \tilde{S}_{k,1} = 1\,,
\tilde{S}_{k2}  =  1 - \eta_{k}/\beta \,,
\nonumber \\
&& \tilde{S}_{k3}  =  -4 + 2\eta_{k}/ \beta \,,
\;
\tilde{S}_{k4}  =  1-\eta_{k}/ \beta \,,
\;
\tilde{S}_{k5} = 1\,,
\;
k = 2,3 \,; \nonumber \\
\quad
&& \tilde{S}_{k3}  =  0\,,
\;
\tilde{S}_{k4} =  -1+\eta_{k}/\beta \,,
\;
\tilde{S}_{k5} = -1\,,
\;
k=4,5\,. \nonumber
\end{eqnarray}
\begin{eqnarray}
&& s_{mn}  =  \frac{\tilde{s}_{mn}}{\sqrt{\sum_{n}\tilde{s}^{2}_{mn}}}\,,
\quad
\tilde{s}_{i1} = -\sum_{n=2}^{5}\tilde{s}_{in}\,,
\nonumber \\
&& \tilde{s}_{i2}  = 1-v_{i}(a_{i}-w_{1}/w_{2}+1)/w_{1}\,,
\;
\tilde{s}_{i3} = 1\,,
\;
\tilde{s}_{i4}  =  1+a_{i} \,,
\nonumber \\
&& \tilde{s}_{i5}  =  1+v_{i}/w_{2} \,,
\;
a_{i} = v_{i}(v_{i}+w_{1}+2 w_{2})/(w_{1} w_{2}) \,.\nonumber
\end{eqnarray}

Energy of the SLS equals
\begin{equation}
E_l \simeq \hbar \omega_{D} {\cal K} \varepsilon_l;
\quad  \varepsilon_l = 3\omega_l^2a^2 + z^2
{\Big (}\frac{1}{2}
 + \frac{z}{3} + \frac{z^2}{4}{\Big )};
\label{eq:el}
\end{equation}
one sees that parameter ${\cal K}$ determines the quantum scale of the SLS
energy. Note that the intensity of phonon emission does not depend on
${\cal K}$ (the unit of $I$ in (\ref{eq:power}) is ${\hbar}\omega_{D}^{2}$).
Consequently, parameter ${\cal K}$ determines the time-scale of energy
relaxation.

We performed calculations of the frequency dependence of the intensity of
phonon emission by the SLS and of the time dependence of the intensity of
phonon emission and the SLS energy (at the temperature $T=0$). The calculation
procedure was the following. First we fixed the initial SLS frequency (usually
at $\omega_{l} = 1.98 \omega_{D}$) and calculated $z$, $a^{2}$, $E_{l}$,
$G(\omega)$, $I$. The second step was to take a lower value of $\omega_{l}$
and respectively calculate new values of parameters. We repeated this
procedure until the frequency of the SLS got value 1.3. The time step was
calculated using the relation $\Delta t = - \Delta E_{l} / I$.

In Fig.~1 the dependence of the intensity of phonon emission on the value of
$\omega_{l}$ is shown. Indeed, the intensity $I$ has peaks
$\sim |\omega_{l} - \omega_{k}|^{-1}$ at critical frequencies $\omega_{k}$
(this was confirmed numerically).
This means that the SLS is unstable in the vicinity of $\omega_{k}$. The time
dependences of the intensity of phonon emission and of the mode energy $E_{l}$
are given in Fig.~2.

In all cases the lowest considered frequency $\omega_{l} = 1.3 \omega_{D}$
corresponds to a small energy as compared to the initial energy for
$\omega_{l}\approx 2\omega_{D}$. Consequently, the main part of the energy of
the SLS with initial frequency $\omega_{l} \sim 2 \omega_{D}$ is lost during
the time considered (note that an extrapolation of $E_{l}(t)$ to larger $t$
also gives finite lifetime of SLS). Depending on the value of $\delta$ and on
the initial energy, this time can be either longer or shorter than that given
by the perturbation theory. The relaxation law is essentially nonexponential:
sharp bursts of phonons are generated at critical values of time $t_k$ when
$\omega_{l}$ approaches $\omega_{k}$ causing relaxation jumps. The time
dependence of the intensity $I$ near the $t_k$ is $I\sim|t-t_k|^{-1/2}$. The
peaks mentioned are caused by the poles in the integrand function
(\ref{eq:power}); they correspond to the emission of quasimonochromatic
phonons and give sharp lines in the spectrum of generated phonons $J(\omega)$
(the integrand function in (\ref{eq:power}) summed over time) in Fig.~3.

The SLS is not fully localized on three central atoms: it always has
wings. Equations (4) and (5) give the following solutions for the amplitudes
$A_{n}$ and the shifts $\xi_{n}$ in the wings:
\vspace{-2mm}
\begin{equation}
A_{n} \simeq (-1)^{n} A_{0} e^{-\mu |n|}\,,
\quad
\xi_{n} \simeq {\rm sign}(n) \frac{\bar{K}_{3} A_{0}^{2}}{2\bar{K}_{2}}
e^{-2\mu|n|}\,,
\quad
n \gg 1
\end{equation}
where $\mu = 2{\rm arch}{\big (}\omega_{l}/\!\omega_{D}{\big )}$.
It can be shown that taking into
account these amplitudes and shifts leads to relaxation jumps of SLSs with
$\omega_l$ close to
$2\omega_{D}$ ($2\omega_{D}-\omega_{l} \stackrel{<}{\sim} 0.01$).
These relaxation jumps are weaker than those caused by the central atoms.

In the case $\delta^{2} < 3/4$ there may exist \cite{kosevich}
large-size SLS with
$\omega_{l}$ close to $\omega_{D}$. The energy of such an SLS
$E_{l} \sim 8\sqrt{2(\omega_{l}-1)}K_{2}^{2}/(K_{4}(3-4\delta^{2}$))
exceeds the unit ($\hbar \omega_{D}$) only if $\delta^{2}$ approaches $3/4$
from below. The problem of quantum decay of this large size SLS is analogous
to that of solitons in continuous media and it will be considered elsewhere.

In conclusion, we examined the quantum relaxation of the odd SLS in a
one-di\-men\-sio\-nal anharmonic monatomic lattice, caused by the zero-point
fluctuations of the phonons and gave an analytical nonperturbative solution of
the problem. We found that the fluctuations can cause generation of very
sharp bursts of quasimonochromatic phonons and very fast, step-wise jumps of the SLS energy. The
rate of energy loss is on average of the order of the phonon quantum
$\hbar \omega_{D}$ per period of vibrations $2\pi/\omega_{D}$. The full
relaxation time is determined by the quantum anharmonicity parameter
${\cal K}$; in chains with potentials corresponding to the quantum crystals,
this time is of the order of a period of vibrations, while in lattices with
potentials of ordinary crystals it may reach thousands of periods.

Supported by the Estonian Science Foundation Grant No. 2274.

\newpage
\noindent
{\Large Figure Captions} \\

\bigskip

\noindent
Figure 1. Frequency dependence of the intensity of emission of phonons by
the odd SLS.The peaks correspond to the regions where the SLS is unstable
withrespect to quantum fluctuations;
$\delta = (K_{3}^{2}/K_{2}K_{4})^{1/2}$ . \\

\bigskip

\noindent
Figure 2. Time dependence of the SLS energy $E_{l}$ (thick dashed line)
and intensity of phonon emission by the SLS $I$ (thick solid line); decay
of the SLS energy calculated with perturbation theory (thin solid line).
Sharp peaks in the $I(t)$ dependence describe emission of phonon bursts by
the SLS. $E_{l} = \varepsilon_{l} K_{2}^{2}/K_{4}$, ${\cal K} =
K_{2}^{2}/(K_{4}\hbar \omega_{D})$. $I$ is measured in units of $\hbar
\omega_{D}^{2}$. \\

\bigskip

\noindent
Figure 3. Spectrum of generated phonons $J(\omega)$ (in a.u.).

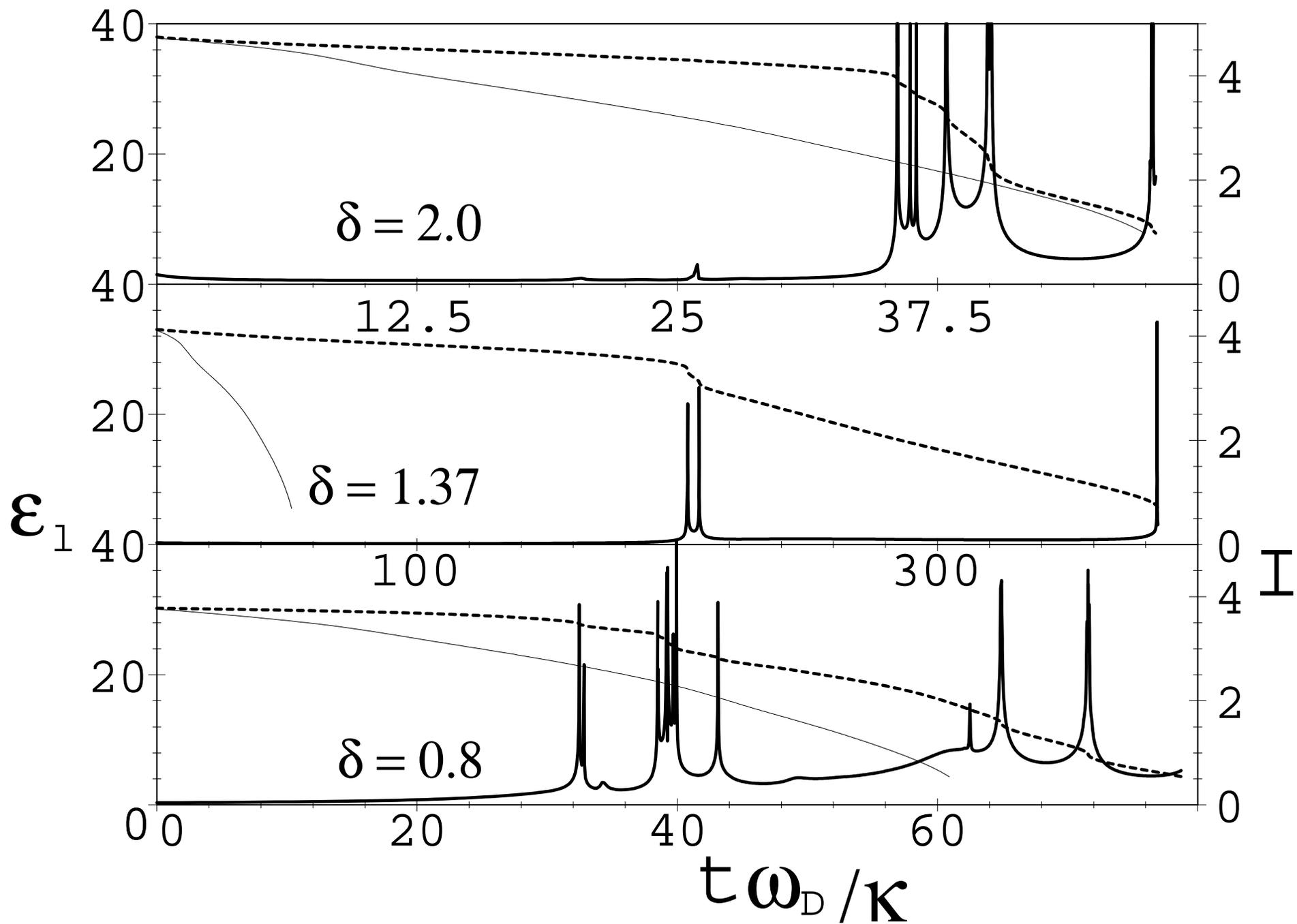
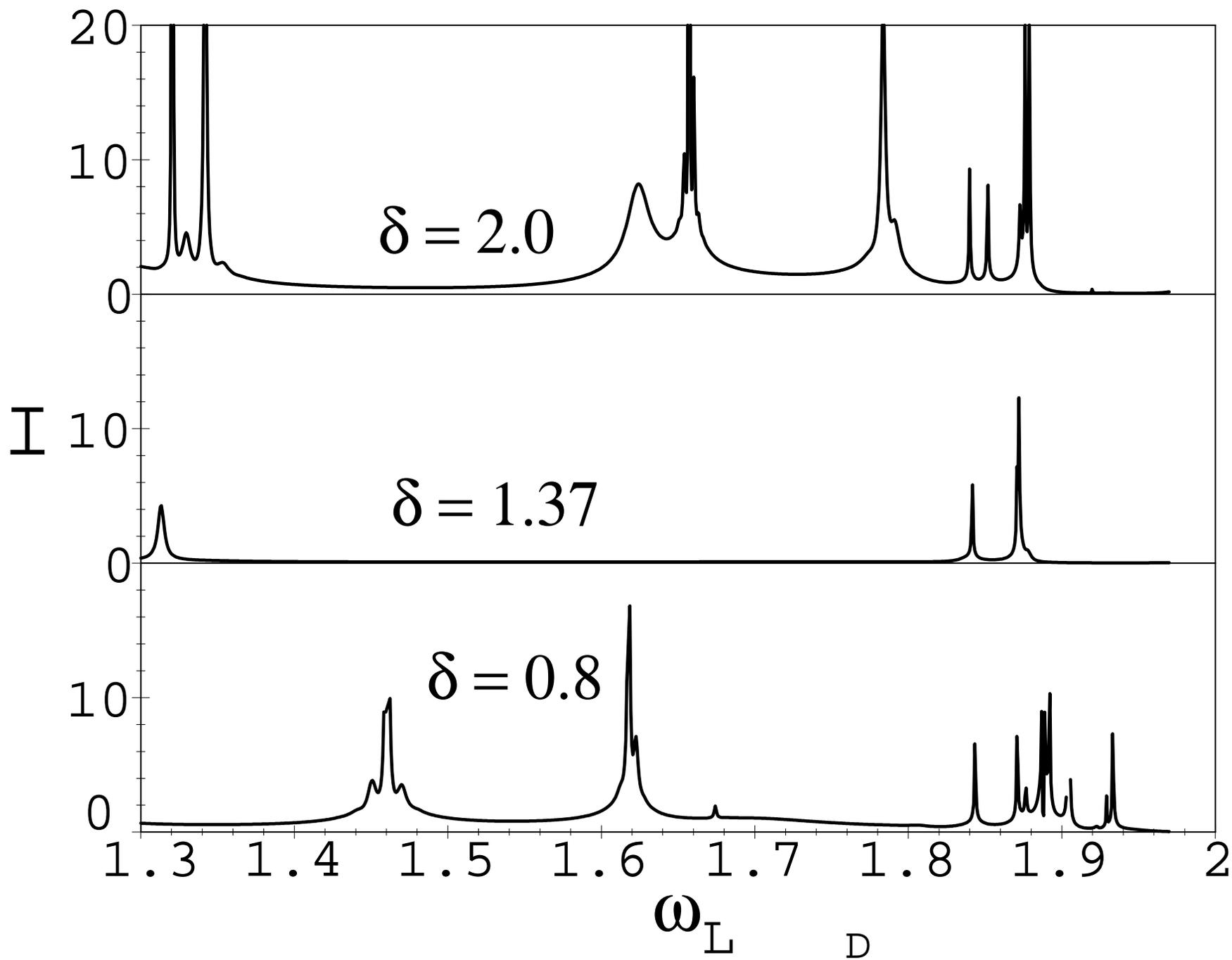
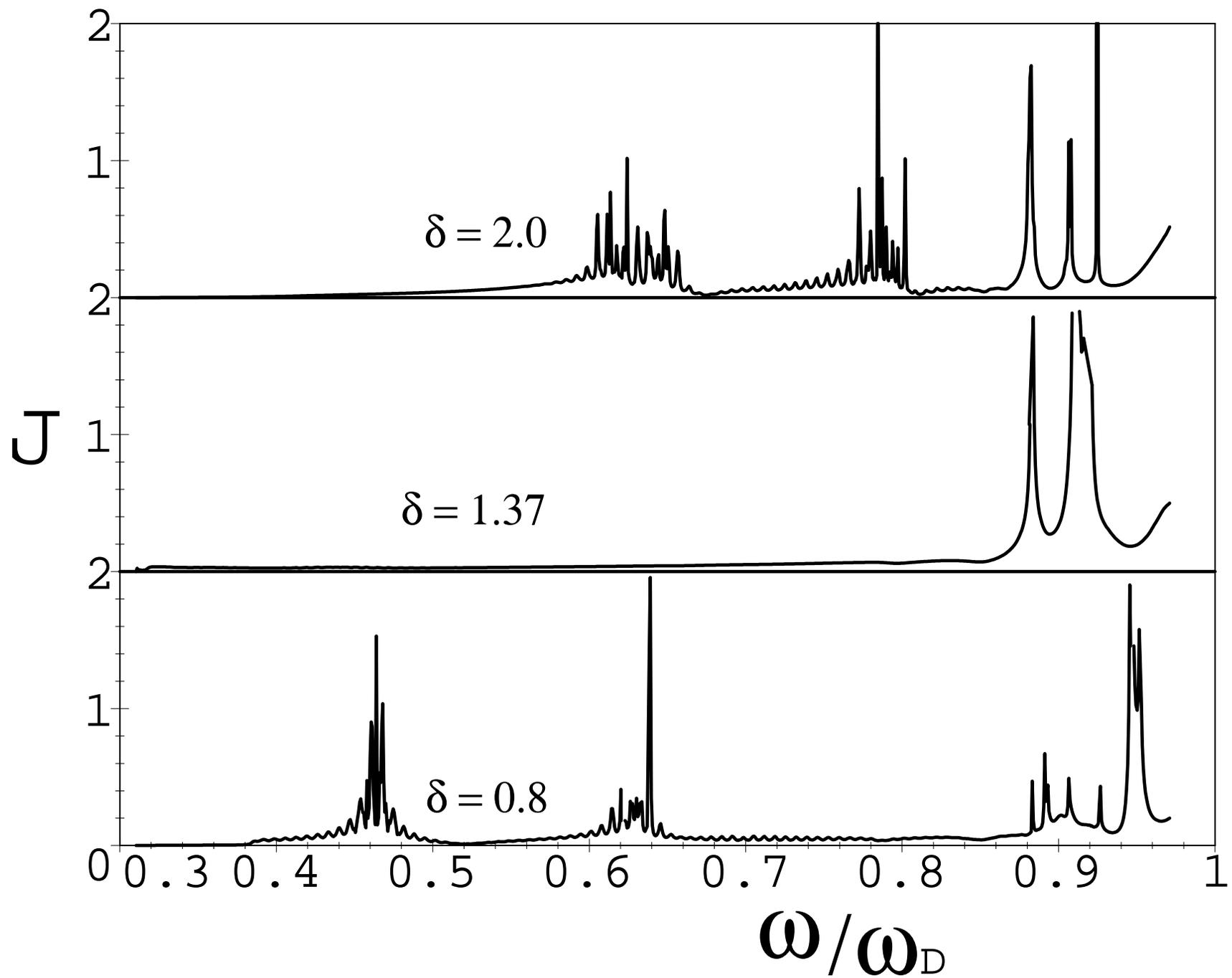
\end{document}